\documentclass[pra,preprint,superscriptaddress,showpacs]{revtex4}
\usepackage{amsmath,amssymb} 
\usepackage{bm}
\usepackage{hyperref}
\usepackage{subfigure}
\usepackage{graphicx}
\usepackage{color}

\begin{document}

\title{Experimental observation of magic-wavelength behavior in optical lattice-trapped $^{87}$Rb}
\author{N.~Lundblad} \email{nlundbla@bates.edu}
\affiliation{Department of Physics \& Astronomy, Bates College, Lewiston, ME 04340, USA}
\affiliation{Joint Quantum Institute, National Institute of Standards and Technology and University of Maryland, 
Gaithersburg, Maryland 20899, USA}

\author{M.~Schlosser}
\affiliation{Institut f\"{u}r Angewandte Physik, Technische Universit\"{a}t Darmstadt, Schlossgartenstra\ss e 7, D-64289 Darmstadt, Germany}

\author{J.~V.~Porto}
\affiliation{Joint Quantum Institute, National Institute of Standards and Technology and University of Maryland, 
Gaithersburg, Maryland 20899, USA}

\date{\today}
\begin{abstract}

We demonstrate the cancellation of the differential ac Stark shift of the microwave hyperfine clock transition in trapped $^{87}$Rb atoms.    Recent progress in metrology exploits so-called ``magic wavelengths," whereby an atomic ensemble can be trapped with laser light whose wavelength is chosen so that both levels of an optical atomic transition experience identical ac Stark shifts.    Similar magic-wavelength techniques are not possible for the microwave hyperfine transitions in the alkalis, due to their simple electronic structure.    We show, however, that ac Stark shift cancellation is indeed achievable for certain values of wavelength, polarization, and magnetic field. The cancellation comes at the expense of a small magnetic-field sensitivity.    The technique demonstrated here has implications for experiments involving the precise control of optically-trapped neutral atoms.

\end{abstract}
\maketitle

Optical trapping is widely used for the manipulation of cold and ultracold neutral atoms.  It is often desirable to optically trap neutral atoms without affecting the internal energy-level spacing of the atoms, particularly in cases (such as in atomic clocks or quantum computing) where the accurate measurement of an atomic transition frequency is paramount.  Unfortunately, for any two internal states of an atom, optical trapping generally induces a {\em differential} shift: this shift presents uncertainty in atomic transition frequencies and also introduces inhomogeneity in the transition, associated with the spatial profile of the trapping potential.  Approaches have been proposed and demonstrated~\cite{JunYe06272008} to minimize these differential shifts in optical transitions, but it is desirable to extend these approaches to other atoms and transitions.  Here we demonstrate a technique (similar to ones recently proposed~\cite{flambaum:220801,Choi:2007p1623}) to eliminate the trapping-light intensity dependence of the ground-state hyperfine transition in alkali atoms.   
 
The differential shift $\delta\nu$ arises from the different frequency-dependent ac polarizabilities of the ground~($g$) and excited~($e$) states of a transition.  Trapping light with frequency $\omega_L$ and electric field ${\bf E}$ shifts the transition frequency by
\begin{equation}
 \delta \nu =\frac{1}{h}\left[ \delta E_e({\bf E},\omega_L)-\delta E_g({\bf E},\omega_L) \right],
\end{equation}
where the light shifts $\delta E_i$ depend on the light polarization through the vector nature of ${\bf E}$. Depending on the polarizability of the the two states, $\delta \nu$ can sometimes be made to vanish (independent of the intensity $ I=\frac{1}{2} c \epsilon_0 |E|^2$) for appropriate choice of $\omega_L$. 

To determine the ways that the differential shift can be cancelled, we consider the light-atom interaction. For dipole-allowed transitions in the low magnetic field limit, where the total angular momentum $F$ is a good quantum number, the light-induced Hamiltonian for states $|n F m_F\rangle$ can be separated into three contributions characterized by the scalar, vector and tensor polarizabilities $\alpha^s$, $\alpha^v$ and $\alpha^t$ \cite{Happer:1967p1572, Deutsch:2009p1873}:
\begin{eqnarray}
{\mathcal H}_{nF}  & = -\left(\frac{1}{2} E_0 \right)^2   & \Biggl[ \biggr. \alpha^s_{nF}
   + \alpha^v_{nF} \left( i \vec{\epsilon}^{\ *} \times \vec{\epsilon}\ \right) \cdot {\bf \hat{F}}  \nonumber \\
 & &  +  \alpha^t_{nF} \left( \frac{ 3 \left| \vec{\epsilon} \cdot {\bf \hat{F}} \right|^2 -  {\bf \hat{F}}^2
   \left| \vec{\epsilon} \right|^2 }{ F (2 F -1)} \right) \biggl. \Biggr],
\end{eqnarray}
where $\hbar {\bf \hat{F}}$ is the total atomic angular moment operator, ${\bf E}=E_0 \vec{ \epsilon}$ and $\vec{\epsilon}$ is a unit-normalized polarization vector. The vector contribution can be treated as arising from an effective magnetic field $\mu_F {\bf B}^{\rm eff}_{nF} =(E_0/2)^2 \alpha^v_{nF} \left( i \vec{\epsilon}^{\ *} \times \vec{\epsilon}\ \right) $, and vanishes for linearly polarized light, where $( i \vec{\epsilon}^{\ *} \times \vec{\epsilon}\ )=0$. The light shift of a given eigenstate $|n F m_F\rangle$ is given by $\delta E_{nFm_F}=\langle n F m_F|{\mathcal H}_{nF} | n F m_F\rangle $. The contributions to the total differential shift $\delta\nu$ from the scalar, vector, and tensor polarizabilities  are $\delta\nu^s$, $\delta\nu^v$, and $\delta\nu^t$, respectively.

If the two states $|e\rangle=|n^\prime F^\prime m_F^\prime\rangle$ and $|g\rangle=|n F m_F\rangle$ have different electronic configurations $n$, the frequency dependence of the polarizabilities can be sufficiently different that a ``magic'' frequency $\omega_*$ may occur even when the light shift is dominated by just the scalar term, so that $\alpha^s_{n^\prime F^\prime}(\omega_*)-\alpha^s_{nF}(\omega_*) = 0$.  Such a cancellation has been found for optical transitions, permitting high accuracy interrogation of narrow optical transitions for optically trapped atoms~\cite{Brusch:2006p1555,Takamoto:2005lr,Boyd:2006p1557}. 

For transitions within the same electronic configuration, such as microwave clock transitions, cancellation is less obvious. The slightly different polarizabilities  have nearly identical frequency dependences, and the light shifts tend to track together as a function of $\omega_L$, without crossing.  (This is true in the far-detuned limit appropriate for optical dipole traps.   In the limit where the detuning is small with respect to hyperfine transitions, cancellation is possible, but at the expense of a substantially increased scattering rate~\cite{Kaplan:2002p1871,Chaudhury:2006p1874}). For example, Rosenbusch et al.~\cite{rosenbusch:013404} showed that for linearly polarized light on alkali atoms, where the vector and tensor contributions to the differential shift are negligible, there is no magic wavelength such that the scalar polarizabilities cancel. (Although for atoms such as aluminum or gallium that have enough internal electronic structure to allow for a tensor contribution, magic wavelengths can be found where the scalar and tensor parts cancel each other \cite{beloy:120801}.)

\begin{figure}[t]
\centering
  \includegraphics[width=\columnwidth]{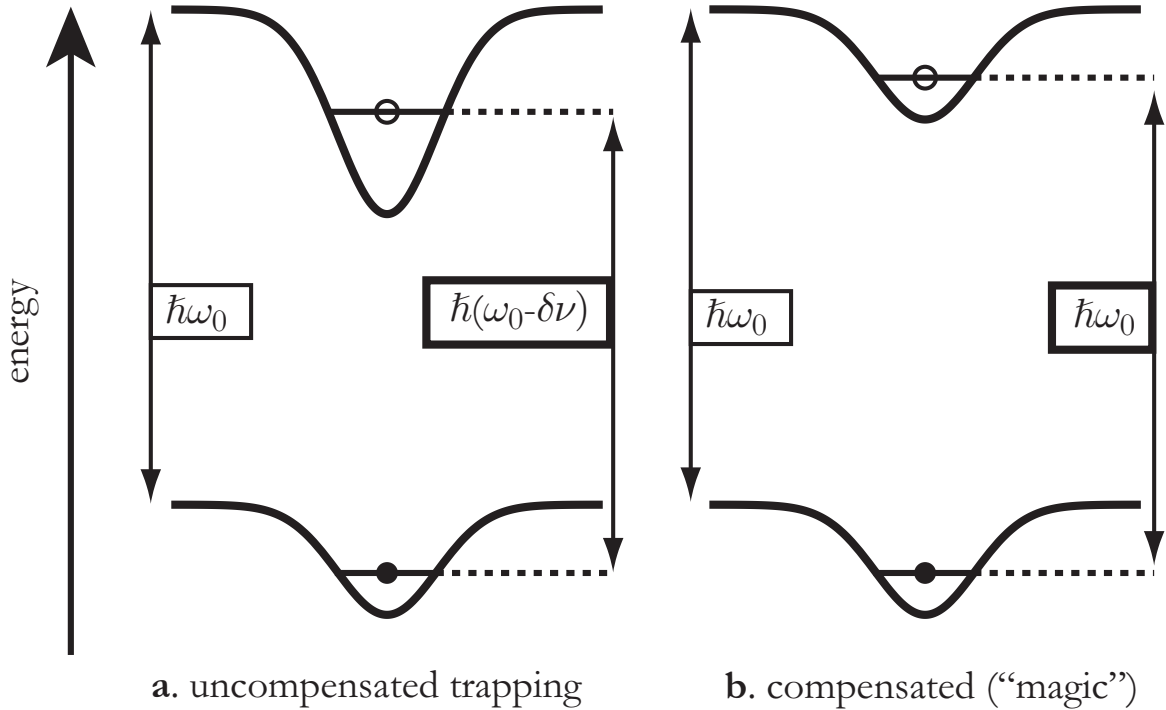}   
    \caption{``Magic wavelength" behavior in optically trapped atoms.       At left: the ground and excited states of an atomic transition experience a differential shift $\delta\nu$ due to frequency-dependent ac polarizabilities.   An ensemble of traps of differing depths leads to inhomogeneous broadening of $\delta\nu$.    At right: an altered trapping configuration with a vanishing $\delta\nu$.    }
\label{scheme}
\end{figure}

Recently there have been proposals~\cite{Choi:2007p1623, flambaum:220801} to use the {\em vector} component of the light shift to cancel the scalar differential shift using elliptically or circularly polarized light on alkali atoms. Since the vector light shift acts like an effective Zeeman field, this approach requires field-sensitive transitions, and depends on the ellipticity of the light.    Additionally, the approach depends on the orientation of the atom determined by the total magnetic field ${\bf B} + {\bf B}^{\rm eff}_{nF}$, where ${\bf B}$ is the external bias field and ${\bf B}^{\rm eff}_{nF}$ points along the local axis of circular polarization. This broadens the discussion to a ``magic surface" in a space of wavelength, polarization and field direction: for most wavelengths there would exist either a magic angle $\theta$ between  ${\bf B}$ and  $ {\bf B}^{\rm eff}_{nF}$ (or fraction of circular polarization) where the differential light shift is cancelled. 

This proposal is challenging, though, due to the fact that the differential vector shift on field sensitive transitions, $\delta \nu^{ v}$, is typically orders of magnitude larger than the differential scalar shift $\delta \nu^{ s}$.   This scale discrepancy can be seen as follows: at lowest order (second order in the light field), the magnitude of the scalar shift $\alpha^s_{nF}$ is independent of $F$. The first non-zero contribution, therefore, to $\delta \nu^s \propto (\alpha^s_{nF^\prime}-\alpha^s_{nF})$ occurs at third order (quadratic in electric field and linear in the hyperfine interaction)~\cite{rosenbusch:013404}. On the other hand, even though the magnitude of the vector shift $|\mu_F {\bf B}^{\rm eff}_{nF}|$ is independent of $F$ at second order, $\delta\nu^v$  acts like a Zeeman shift, depending on the projections $m_F$, $m_F^\prime$.  It results from fine structure coupling~\cite{Happer:1967p1572,Deutsch:2009p1873} and has significant non-zero contribution at second order. In heavy atoms like rubidium or cesium, the fine structure is orders of magnitude larger than the hyperfine structure, and $\delta\nu^v \gg \delta\nu^s$. Cancellation requires an accurate rescaling of the size of $\delta \nu^{ v}$, either by controlling a small amount of circular polarization at the level of $10^{-6}$~\cite{Choi:2007p1623}, or tuning the angle $\theta$ between ${\bf B}$ and ${\bf B}^{\rm eff}$ such that its cosine is precisely set to a value smaller than 10$^{-2}$\cite{flambaum:220801}.




In this paper, we show that by using $|F, m_F=0\rangle$ states at {\it nonzero magnetic field} we can achieve the light shift cancellation without sensitive control of the polarization $\vec{\epsilon}$ or the angle $\theta$. Rather than relying upon a careful projection between two vectors, we use the small field sensitivity experienced at nonzero magnetic field by atoms in the $m_F=0$ ground states.  The residual vector shift is the response of the transition frequency to a perturbation $B^{\rm eff}$ of the bias field $B$.     We parametrize the field sensitivity of the transition (which arises from the local slope of the so-called quadratic Zeeman shift) using an effective magnetic moment $\mu^\prime/h=d\nu/dB$, a parameter which is easier to precisely control at small values  (via the external bias field B) than $\vec{\epsilon}$ or $\cos{\theta}$.       By applying this small vector light shift on individual $^{87}$Rb atoms trapped in sites of an optical lattice, we show that we can eliminate the dependence of the clock transition frequency on the intensity of the trapping light,  achieving the goal of magic-wavelength behavior in an alkali-metal microwave transition.    For details, see theoretical work elsewhere~\cite{Derevianko:2009p1922}.

One can estimate the field sensitivity required for alkali atoms in purely circularly polarized light by evaluating the ratio of the scalar differential shift $\delta \nu^s$ to the differential vector shift $\delta \nu^v$.  Ignoring contributions from excited states other than the lowest $P$ levels,  $\delta \nu^s$ is determined by the slightly different laser detuning from the excited state due to the $^{2}S_{1/2}$ ground state hyperfine splitting $\Delta_{\rm HF}$:
\begin{equation}\label{scaling}
\delta \nu^s \propto \left(\frac{1}{\Delta+\Delta_{\rm HF}} -\frac{1}{\Delta}\right) \simeq -\frac{1}{\Delta}\left(\frac{\Delta_{\rm HF}}{\Delta} \right),
\end{equation}
where $\Delta$ is the laser detuning from the $P$ states.   Given $\delta E \propto 1/\Delta$,  $\delta\nu^s$ therefore scales as $\delta \nu^s \simeq \delta E (\Delta_{\rm HF}/\Delta)$.  The vector light shift, however, depends on the {\em fine} structure splitting $\Delta_{\rm F}$ of the excited $P$ states~\cite{Deutsch:1998p1335}. For a hypothetical field-sensitive transition with sensitivity $\mu_B$, a similar argument as for Eq.~\ref{scaling} shows that  $\delta \nu^v \propto (1/3 \Delta)(\Delta_{\rm F}/\Delta)$, so that $\delta \nu^v \simeq (1/3) \delta E (\Delta_{\rm F}/\Delta)$. The required field sensitivity $\mu^\prime/\mu_B$ to allow for cancellation is therefore given approximately by  $\delta \nu^s/ \delta \nu^v = 3 \Delta_{\rm HF}/ \Delta_{\rm F}$, which is independent of detuning. For $^{87}$Rb, $3 \Delta_{\rm HF}/ \Delta_{\rm F}= 0.0029$, giving a required field sensitivity of $0.0029\mu_B/h = 40$~kHz/mT. For cesium, the required sensitivity is 23~kHz/mT. This points to a limitation of this approach: obtaining insensitivity to light intensity requires a sensitivity to magnetic fields. 

Our experiment uses an ensemble of single atoms trapped at the sites of a 3D optical lattice, created by adiabatically loading  a Bose-Einstein condensate (BEC) of $\lesssim$~10$^5$ $^{87}$Rb atoms into a lattice generated by a Ti:sapphire laser operating at $\lambda=811.5$ nm.   A linearly polarized ``vertical" lattice along $\hat{z}$ divides the BEC into a stack of independent 2D condensates, and a separate, deformable lattice created by beams along $\hat{x}$ and $\hat{y}$ completes the 3D lattice~\cite{sebby-strabley:033605}.     This $xy$ lattice can be transformed from a standard  linearly polarized $\lambda/2$-periodic lattice into a double-well configuration, where both the lattice intensity and polarization can differ significantly between ``left" and ``right" sides of the double well. The details of this lattice are described extensively elsewhere~\cite{sebby-strabley:033605,lee:020402}.   The applied bias field ${\bf B}$ lies in the $xy$ plane.

\begin{figure}[t!]
\centering
  \includegraphics[width=\columnwidth]{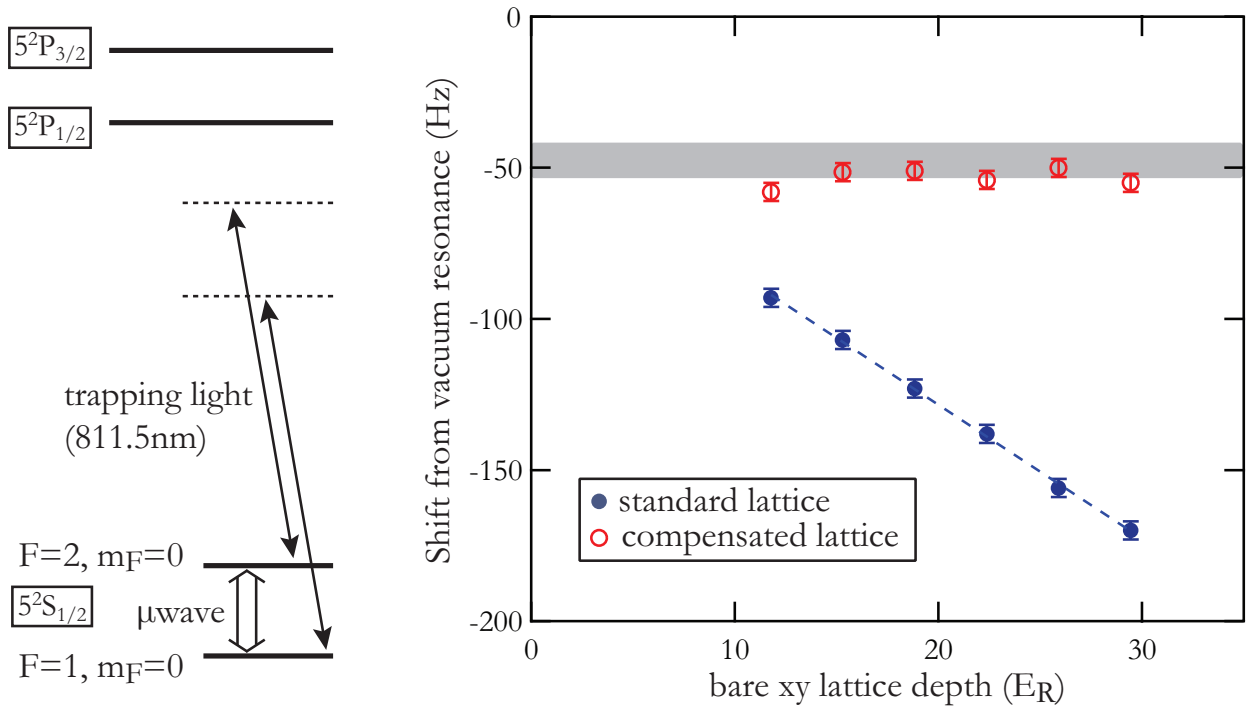}
      \caption{At left, an energy-level diagram of our system, including trapping light and the hyperfine microwave transition.   At right, the measured differential shift $\delta\nu$ of the $^{87}$Rb $|1,0\rangle\rightarrow |2,0\rangle$ transition as a function of the $xy$ lattice intensity, with a fixed vertical-lattice depth of 21 $E_R$.   ($E_R=h^2 /2m\lambda^2$ is the single-photon recoil energy.)    We observe $\delta\nu$ in the linearly polarized lattice (closed circles) and in the compensated (elliptically polarized) lattice (open circles) as a function of $xy$-lattice intensity, using Ramsey interferometry.     We parametrize  the intensity used in both configurations by the equivalent depth (in $E_R$) of the linearly-polarized lattice at a given laser intensity.      In the compensated lattice, $\delta\nu$ is essentially independent of intensity.      The measured differential shift due to the vertical lattice (including uncertainty) is indicated by the gray band.    The depth of the lattices were calibrated using pulsed matter-wave diffraction. }    
\label{leftright}
\end{figure}

We use the $5 ^2 S_{1/2}$ $|1,0\rangle\rightarrow|2,0\rangle$ transition in $^{87}$Rb, which at zero field lies near 6.834 682 611 GHz, and increases with field approximately as 57.5 kHz/(mT)$^2$.
  To measure the differential shift, we perform microwave Ramsey interferometry on the ensemble of atoms using $\pi/2$-pulses and a variable Ramsey delay, with the pulses detuned from free-space resonance (as calculated by the Breit-Rabi formula~\cite{Breit:1931p1559}) by 1 kHz.   The observed Ramsey fringe shifts from free-space resonance as a function of light intensity, yielding a $> 150$ Hz shift at high intensity (lattice depth $\simeq$ 30 $E_R$).  This is the shift we seek to eliminate.   

In order to cancel the differential shift, we transform the $xy$ lattice into a configuration placing an effective magnetic field ${\bf B}^{\rm eff}$ (originating from an increased local elliptical polarization) on every other lattice site, where ${\bf B}^{\rm eff}$  is aligned parallel to ${\bf B}$. After Ramsey interferometry, we record only the populations of atoms on the effective-field sites, discarding atoms on the unused sites through a ``dumping" process~\cite{sebby-strabley:200405,lee:020402}.  In Fig.~1 we show the measured differential shift of the $|1,0\rangle\rightarrow |2,0\rangle$ transition as a function of the $xy$ lattice intensity, both in the linearly polarized lattice and in the elliptically polarized lattice.   
The latter clearly shows a strongly diminished dependence of $\omega_L$ on $xy$ lattice-beam intensity.     
The remaining $\simeq 50$~Hz shift from bare vacuum resonance is due to the differential light shift caused by the confinement in the linearly polarized vertical lattice.    We measured this shift using Ramsey interferometry and varying the intensity of the vertical lattice, yielding an offset value of 48(4) Hz\footnote{Unless otherwise stated, all uncertainties herein reflect the uncorrelated combination of 1$\sigma$ statistical and systematic uncertainties.}.   We believe it should be possible to cancel this shift as well, using a bias field with some component along $z$ as well as in the $xy$ plane. We observed compensation at a bias field $B$ of 0.4314(3) mT, corresponding to $\mu^\prime/h=49.6$~kHz/mT.  From a model of the degree of ellipticity in our lattice, we estimate that ${\bf B}^{\rm eff}$ is only 0.77(4) times what it would be for purely circularly polarized light, requiring a scaling of our above estimate of  $\mu^\prime/h$ from 40 kHz/mT to 52(3) kHz/mT, in agreement with observation.   

This work has implications for experiments requiring the accurate control of trapped neutral atoms. In particular, the use of optically trapped neutral atoms for quantum information applications can be adversely affected by fluctuations and inhomogeneity  in the trapping and control lasers~\cite{Lundblad:2009p1492}.    The strong reduction of the differential shift presented in this work offers a possible solution to these issues.      More generally, this work shows that differential insensitivity to light can be obtained at the cost of slightly increased sensitivity to magnetic fields; it may be useful in situations where the light field fluctuations and inhomogeneity are more difficult to control than the global magnetic field.   

\begin{acknowledgments}
We thank Andrei Derevianko, Ivan Deutsch, Saijun Wu, and Stephen Maxwell for helpful discussion.  This work was partially supported by IARPA and ONR.  NL acknowledges support from the NRC. 
\end{acknowledgments}

\bibliographystyle{apsrevNOURL}


\end{document}